\documentclass[lettersize,journal]{IEEEtran}
\usepackage{amsmath,amsfonts}
\usepackage{amssymb}
\usepackage[noend]{algpseudocode}
\usepackage{algorithmicx}
\usepackage{algorithm}
\usepackage{array}
\usepackage[caption=false,font=normalsize,labelfont=sf,textfont=sf]{subfig}
\usepackage{textcomp}
\usepackage{stfloats}
\usepackage{url}
\usepackage{verbatim}
\usepackage{graphicx}
\usepackage{cite}
\usepackage{cuted}
\usepackage{color}
\usepackage{multicol}
\newcommand{\ket}[1]{\left|#1 \right\rangle}

\newcommand{\ketbra}[2]{\left| #1 \rangle \langle #2 \right|}

\begin{document}

\title{Experimental implementation of quantum algorithm for association rules mining}

\author{Chao-Hua Yu
\thanks{Chao-Hua Yu is with the School of Information Management, Jiangxi University of Finance and Economics, Nanchang 330032, China (quantum.ych@gmail.com).}
}

\markboth{IEEE JOURNAL ON EMERGING AND SELECTED TOPICS IN CIRCUITS AND SYSTEMS \LaTeX\ Class Files,~Vol.~14, No.~8, August~2022}%
{Shell \MakeLowercase{\textit{et al.}}: A Sample Article Using IEEEtran.cls for IEEE Journals}


\maketitle

\begin{abstract}

Recently, a quantum algorithm for a fundamentally important task in data mining, association rules mining (ARM), called qARM for short, has been proposed. Notably, qARM achieves significant speedup over its classical counterpart for implementing the main task of ARM, i.e., finding frequent itemsets from a transaction database. In this paper, we experimentally implement qARM on both real quantum computers and a quantum computing simulator via the IBM quantum computing platform. In the first place, we design quantum circuits of qARM for a 2$\times$2 transaction database (i.e., a transaction database involving two transactions and two items), and run it on four real five-qubit IBM quantum computers as well as on the simulator. For a larger 4$\times$4 transaction database which would lead to circuits with more qubits and a higher depth than the currently accessible IBM real quantum devices can handle, we also construct the quantum circuits of qARM and execute them on ``aer\_simulator'' alone. Both experimental results show that all the frequent itemsets from the two transaction databases are successfully derived as desired, demonstrating the correctness and feasibility of qARM. Our work may serve as a benchmarking, and provide prototypes for implementing qARM for larger transaction databases on both noisy intermediate-scale quantum devices and universal fault-tolerant quantum computers.
\end{abstract}


\section{Introduction}
\IEEEPARstart{Q}{uantum} computing takes advantages of quantum mechanical principles, such as quantum superposition and quantum entanglement, to implement computational tasks \cite{QCQI10}. Its most notable advantage over the classical computing lies in the fact that quantum computing has the potential to solve certain problems more efficiently. Up to now, a number of quantum algorithms addressing various problems have been proposed, most of which fall into four categories. The first one aims at simulating behaviors of quantum systems---the original motivation of quantum computing \cite{Feynman82}, which is well known as Hamiltonian simulation, and exponential speedup can be achieved under certain conditions \cite{CMNetal18}. The second one is embodied by Shor's quantum algorithm for integer factoring \cite{Shor94} and exponential speedup can be achieved. Shor's algorithm poses a serious threat to the RSA-based cryptography systems that are now widely used in daily life. The third one focuses on the searching problem, for which the first quantum algorithm was proposed by Grover \cite{Grover97} and its generalized version is known as amplitude amplification \cite{AA02}. The last one builds on the quantum algorithm for solving linear systems of equations proposed by Harrow, Hassidim and Lloyd (HHL algorithm) \cite{HHL09} and mainly solves linear algebra problems arising from machine learning and data mining. To narrow the gap between theory and practicality, experimental realizations of Hamiltonian simulation, Shor's algorithm, Grover's quantum searching algorithm and HHL algorithm on different types of quantum computers were conducted in the refs.\cite{BCC06,YWXD06}, the refs.\cite{VSBetal01,LBYP07,LWLetal07,MLLetal12}, the refs.\cite{LYLetal01,GFB17} and the refs.\cite{CWSetal13,PCYetal14}, respectively.  

In light of HHL algorithm proposed in 2009, quantum computing has been brought into the fields of machine learning and data mining for better computational performance in problem-solving, which becomes an emerging research area with worldwide attention. As of now, a number of quantum algorithms have been proposed to implement various tasks in machine learning and data mining, mainly including data classification \cite{LMR13,PL13,QSVM14,CD16,SFP17,SP18,SK19,HCTetal19}, regression analysis \cite{WBL12,SSP16,Wang17,LZ17,YGLetal19,YGW21}, clustering analysis \cite{LMR13,ABG13,AP21,WSZ22}, dimensionality reduction \cite{CD16,QPCA14,YGLW19,DYXL19,LSLL20,SDPC21}, anomaly detection \cite{LP18,LSLL19,KMFB21,UBGG22}, and association rules mining \cite{QARM16}. Fortunately, it has been shown that most of these quantum algorithms are able to achieve significant speedup over their classical counterparts under certain conditions; for example, the quantum support vector machine is able to classify a new data exponentially faster than the classical support vector machine, if efficient quantum access to the training dataset is available and the training data matrix has a low rank \cite{QSVM14}. Overviews on quantum algorithms for machine learning and data mining can be seen in the refs. \cite{QDM14, QML17, DB18}. However, few of these quantum algorithms have ever been experimentally demonstrated, and the rare existing related experimental implementations mainly focus on the supervised tasks, such as data classification \cite{LLXD15,CWSetal15}.

 In this paper, we aim for a quantum algorithm for a fundamentally important unsupervised task of association rules mining (ARM), named qARM for short, and explore how and how far we can bring it to reality from theory. Specifically, we experimentally implement qARM on both real IBM quantum computers and a noise-free IBM quantum simulator ``aer\_simulator''. Based on a basic quantum oracle of accessing the database, qARM is able to bring off the main task of ARM, i.e., finding out all the frequent itemsets from a transaction database, with significant speedup over the classical ARM algorithm. Our experimental implementation of qARM consists of two parts. In the first one, we work for an engineered 2$\times$2 transaction database (i.e., a transaction database involving two transactions and two items),  design the quantum circuit following qARM, and run it on four real five-qubit IBM quantum computers (``ibmq\_quito", ``ibmq\_manila", ``ibmq\_lima" and ``ibmq\_nairobi") and on the noise-free IBM quantum simulator ``aer\_simulator''. The second one aims for a larger engineered 4$\times$4 transaction database, which gives rise to qARM circuits with more qubits and a higher depth than currently accessible IBM real quantum devices can handle, we also construct the corresponding quantum circuits, and execute them on ``aer\_simulator'' alone. The experimental results of the two parts demonstrate the correctness and feasibility of qARM on quantum computers. Moreover, our implementation may benchmark the implementation of qARM and provide prototypes for more implementations of qARM for larger transaction databases. 

The rest of this paper is organized as follows. In section ~\ref{sec:qARMrev}, we briefly review the qARM algorithm. Then, in section ~\ref{sec:experiments}, experimental implementation of qARM for two engineered transaction datasets are conducted and the experimental results are analyzed. Conclusions are drawn in the last section.

\section{Quantum algorithm for association rules mining: a review}\label{sec:qARMrev}

We first recall the concepts and the classical algorithmic approach for ARM, and then review how this task can be efficiently implemented via qARM.

\subsection{Classical ARM}
\begin{figure}[!h]
        \centering
        \includegraphics[width=3in]{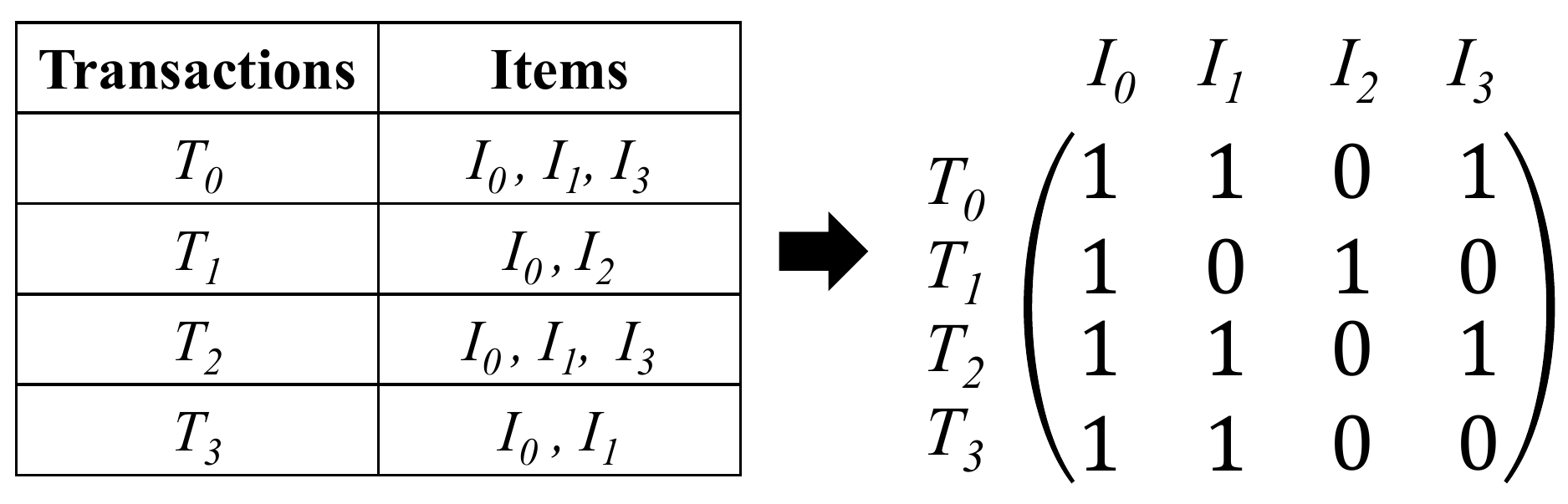}
        \caption{Left: a transaction example with four transactions $\{T_0,T_1,T_2,T_3\}$ and four items $\{I_0,I_1,I_2,I_3\}$. Right: the binary matrix representation of the database. For a real-life instance, the four transactions could be food four different customers have bought from a supermarket, and the four items could be the food of bread, cheese, butter and milk. }
        \label{fig:TDExample}
\end{figure}

 Given a transaction database $\mathcal{T}=\{T_0,T_1,\ldots, T_{N-1}\}$ that involves $N$ transactions  and $M$ items with each transaction being a subset of the $M$-item set $\mathcal{I}=\{I_0,I_1,\ldots,I_{M-1}\}$, i.e, $T_i \subseteq \mathcal{I}$ for $i=0,1,\ldots,N-1$, which can also be described by a binary matrix $D$ with $D_{ij}=1$ if $I_j\subseteq T_i$ and $D_{ij}=0$ otherwise (for $i=0,1,\ldots,N-1$ and $j=0,1,\ldots,M-1$), ARM mainly aims to find out all the \textit{frequent} itemsets (an itemset is a set of items) whose support is greater than or equal to a predetermined threshold $s_{min}$. For simplicity, we hereafter assume $N$ and $M$ are powers of two; if they are not, we extend the database by inserting zeros into $D$. Here, the support of an itemset $X$ is defined as the percentage of the transactions that contains $X$, i.e.,
\begin{align}
{\rm supp}(X)=\frac{|\{T_i|X\subseteq T_i, i=0,1,\ldots, N-1\}|}{N}.
\end{align}
For example, a transaction database with four transactions and  four items and its binary matrix representation are given in Fig. \ref{fig:TDExample}. If we set $s_{min}=0.5$, the frequent itemsets of this database would be $\{I_0\}$, $\{I_1\}$, $\{I_3\}$, $\{I_0,I_1\}$ and $\{I_1,I_3\}$. 

The most famous algorithm for achieving the above goal, i.e., mining all the frequent itemsets from a transaction dataset, is referred to as \textit{Apriori} algorihm. It works iteratively and in the $k$th iteration, two procedures are carried out as follows.
\begin{itemize}[]
\item (P1) Calculate the supports of all the \textit{candidate} $k$-itemsets (a $k$-itemset is a set of $k$ items) denoted by $\mathcal{C}^{(k)}$, which can be obtained by the procedure (P2) of the previous iteration when $k>1$ or is just all the items of the transaction database $\mathcal{C}^{(k)}=\{\{I_0\},\{I_1\},\ldots,\{I_{M-1}\}\}$ when $k=1$, and find out the candidate  $k$-itemsets with support $\ge s_{min}$, forming the frequent $k$-itemsets $\mathcal{F}^{(k)}$  with associated supports $S^{(k)}$.

\item (P2) Generate the candidate $(k+1)$-itemsets $\mathcal{C}^{(k+1)}$ from $\mathcal{F}^{(k)}$ via two steps: join step and prune step.
\end{itemize}
It is worth noting that in each iteration the procedure (P1) is much more time-consuming than the procedure (P2), since (P1) requires scanning the whole transaction database for computing supports, while no access to the database is needed in (P2). The whole process of Apriori algorithm for mining frequent itemsets is described in Algorithm \ref{alg:cARM}, where lines 4-10 correspond to the procedure (P1) and line 11 corresponds to the procedure (P2).
\begin{algorithm}[!htb]
        \caption{Apriori algorithm for mining frequent itemsets} 
        \label{alg:cARM}
        \hspace*{0.02in} {\bf Input:}  $\mathcal{T}=\{T_0,T_1,\ldots T_{N-1}\}$, $\mathcal{I}=\{I_0,I_1,\ldots I_{M-1}\}$, and the minimum support threshold $s_{min}$.\\
        \hspace*{0.02in} {\bf Output:} All the frequent itemsets and their supports.
        \begin{algorithmic}[1]
        \State $k$=1
        \State $\mathcal{C}^{(k)}=\{\{I_0\},\{I_1\},\ldots,\{I_{M-1}\}\}$
        \While{$\mathcal{C}^{(k)} \neq \emptyset$}
                \State $\mathcal{F}^{(k)}=\{\}$
                \State $S^{(k)}=\{\}$
                \For{$X$ in $\mathcal{C}^{(k)}$}
                        \State Calculate ${\rm supp}(X)$ by scanning $\mathcal{T}$
                        \If{${\rm supp}(X) \ge s_{min}$} 
                                \State plug $X$ into $\mathcal{F}^{(k)}$
                                \State plug ${\rm supp}(X)$ into $S^{(k)}$
                        \EndIf       
                \EndFor
                \State Generate $\mathcal{C}^{(k+1)}$ from $\mathcal{F}^{(k)}$ via join and prune 
                \State $k=k+1$
        \EndWhile
        \State \Return $\cup_{k}\mathcal{F}^{(k)}$ and $\cup_{k}S^{(k)}$
        \end{algorithmic}
\end{algorithm}

\subsection{qARM}
\label{subsec:qARM}
The qARM algorithm \cite{QARM16} also works in an iterative way as the classical ARM, but for each iteration it focuses on speeding up the procedure (P1) on a quantum computer and leaves the procedure (P2) on a classical computer, because the procedure (P1) as mentioned above plays a dominant role in each iteration in terms of time cost. The basic idea of qARM is that, in the procedure (P1) of each iteration, estimating the supports of all the candidate itemsets in fact can be done via amplitude estimation \cite{AA02}  in quantum parallel \cite{QARM16} and finding out candidate itemsets whose supports are greater than or equal to $s_{min}$ can implemented via amplitude amplification \cite{AA02}. Execution of the qARM algorithm is based on the basic quantum oracle denoted by $O_B$ that accesses the whole database $\mathcal{T}$ represented by the binary matrix $D$ and acts on three quantum registers as 
\begin{align}
O_B\ket{i}\ket{j}\ket{a}=\ket{i}\ket{j}\ket{a\oplus D_{ij}},
\label{eq:OB}
\end{align}
for $i=0,1,\ldots,N-1$ and $j=0,1,\ldots,M-1$.
Here all the three registers are in the quantum computational states: the first one stores the indices of transactions, the second one stores the  indices of items, and the last one is just an ancillary qubit.  Efficient synthesizing the quantum circuit of $O_B$ can be achieved by Gray-code-inspired techniques \cite{AAM19,SBBetal21}.  Setting $\ket{a}=(\ket{0}-\ket{1})/\sqrt{2}$, we can use one $O_B$ to readily implement the quantum oracle $O^{(1)}$ for each $1$-itemset $\{I_j\}$: 
\begin{align}
O^{(1)}\ket{i}\ket{j}=(-1)^{D_{ij}}\ket{i}\ket{j},
\label{eq:O1}
\end{align}
where the ancillary qubit is discarded for no state change in the operation. This quantum oracle tells us whether $I_j \subseteq T_i$ for any $1$-itemset $\{I_j\}$, working as the quantum oracle in the standard Grover's quantum search algorithm \cite{Grover97}. 
Moreover, $2k$ basic oracles $O_B$ together with $\Theta(k)$ one-qubit and two-qubit gates are utilized to implement a more complicated quantum oracle $O^{(k)}$ , which for each $k$-itemset $X^{(k)}=\{I_{j1},I_{j2},\ldots,I_{jk}\}$ ($j_1,j_2,\ldots,j_k \in \{0,1,\ldots,M-1\}$) acts as 
\begin{align}
\label{eq:Ok}
        O^{(k)}\ket{i}(\ket{j_1}\ket{j_2}\ldots\ket{j_k})=(-1)^{\tau(i,X^{(k)})}\ket{i}(\ket{j_1}\ket{j_2}\ldots\ket{j_k}),
\end{align}
where $\tau(i,X^{(k)})=\prod_{l=1}^kD_{ij_l}$. The quantum oracle $O^{(k)}$ identifies whether $X^{(k)} \subseteq T_i$ for any $k$-itemset $X^{(k)}$; if $X^{(k)} \subseteq T_i$, the phase would be flipped, otherwise nothing happens.  Then we can further construct the Grover operation $G^{(k)}$ defined as 
\begin{align}
\label{eq:Gk}
        G^{(k)}=\left[(2\ketbra{\chi_N}{\chi_N}-\mathbb{I}_t)\otimes \mathbb{I}_i\right]O^{(k)},
\end{align}
where $\ket{\chi}_N:=\sum_{i=0}^{N-1}\ket{i}/\sqrt{N}$, $\mathbb{I}_t$ and $\mathbb{I}_i$ are identity operation on the first quantum register of $\ket{i}$ storing transaction indices, and on the second register of  $\ket{j_1}\ket{j_2}\ldots\ket{j_k}$ storing the item indices of any $k$-itemset $X^{(k)}$. Confining the second register to the state $\ket{j_1}\ket{j_2}\ldots\ket{j_k}$ for the specific $k$-itemset $X^{(k)}$, $G^{(k)}$ has two eigenvalues denoted by $\pm e^{-2\iota \theta^{(k)}}$ ($\iota=\sqrt{-1}$) and two corresponding eigenvectors denoted by $\ket{\phi^{(k)}_\pm}$, and the support of $X^{(k)}$ can be derived by 
\begin{align}
{\rm supp}(X^{(k)})=\sin^2(\theta^{(k)}).
\end{align}
This means that one can estimate ${\rm supp}(X^{(k)})$ by estimating $\theta^{(k)}$ using amplitude estimation technique with the Grover operation $G^{(k)}$ \cite{AA02}.

The whole qARM algorithm is described in Algorithm \ref{alg:qARM} and the $k$th iteration for mining frequent $k$-itemsets can be summarized by the following four steps.

(Step 1) Prepare the state of three quantum registers  with qubits
\begin{align}
\label{eq:StartingState}
        \left(\frac{\sum_{t=0}^{T-1}\ket{t}}{\sqrt{T}}\right) \ket{\chi_N} \left(\frac{\sum_{j=1}^{M_c^{(k)}}\ket{C_j^{(k)}}}{\sqrt{M_c^{(k)}}}\right).
\end{align}
Here the state of the last quantum register stores all $M_c^{(k)}$ candidate $k$-itemsets $\mathcal{C}^{(k)}=\{C_j^{(k)}|j=1,2,\ldots,M_c^{(k)}\}$ in quantum superposition; for example, if $\mathcal{C}^{(2)}=\{\{I_0,I_1\},\{I_0,I_2\},\{I_1,I_2\}\}$ with $k=2$ and $M_c^{(2)}=3$, the state would be
\begin{align*}
        \frac{\ket{00}\ket{01}+\ket{00}\ket{10}+\ket{01}\ket{10}}{\sqrt{3}}.
\end{align*}
The state can be generated by performing a unitary operation on a number of qubits $\ket{0\ldots 0}$; the unitary operation could be just Hadamard operations on all qubits when $k=1$, or can be implemented by a low-depth quantum circuit by introducing ancillary qubits \cite{STYetal21}. It is notable that, according to the above analysis for any $k$-itemset, the Grover operation $G^{(k)}$ would have two eigenvalues denoted by $\pm e^{-2\iota \theta_j^{(k)}}$ and two associated two eigenvectors denoted by $\ket{\phi_{j\pm}^{(k)}}$ when the state of the third quantum register $G^{(k)}$ acts on is confined to $\ket{C_j^{(k)}}$, and the relationship between the support the $j$th candidate $k$-itemset $C_j^{(k)}$ denoted by $s_j^{(k)}$ and the angle $\theta_j^{(k)}$ can be described as 
\begin{align}
\label{eq:sjk}
s_j^{(k)}=\sin^2(\theta_j^{(k)}).
\end{align}
The state of the first two quantum registers can be readily generated by a bunch of Hadamard operations on qubits with zero state.

(step 2) Performing the amplitude estimation of the Grover operation $G^{(k)}$ on the three registers, we have 
\begin{multline}
\label{eq:pae}
       \sum_{j=1}^{M_c^{(k)}} \frac{1} {\sqrt{2M_c^{(k)}}}\Biggl(e^{\iota\theta_j^{(k)}}\ket{\frac{\theta_j^{(k)}}{\pi}}\ket{\phi_{j+}^{(k)}}- \\
e^{-\iota\theta_j^{(k)}}\ket{1-\frac{\theta_j^{(k)}}{\pi}}\ket{\phi_{j-}^{(k)}}\Biggr) \ket{C_j^{(k)}} ,
\end{multline} 
where the states of the first quantum register $\ket{\theta_j^{(k)}/\pi}$ (or $\ket{1-\theta_j^{(k)}/\pi}$) encode the supports of all the candidate $k$-itemset $C_j^{(k)}$ stored in the third register $\ket{C_j^{(k)}}$ (i.e. $s_j^{(k)}$) in quantum parallel, according to Eq.~\eqref{eq:sjk}.

(Step 3) Marking the state of the first register for $s_j^{(k)}=\sin^2(\theta_j^{(k)})\ge s_{min}$, we perform amplitude amplification \cite{AA02} to search out a part of the state of Eq.~\eqref{eq:pae} with frequent candidate $k$-itemset.

(Step 4) Measure the  last and the first register to read out the classical information of frequent $k$-itemsets and their associated supports  respectively.

\begin{algorithm}[!htb]
        \caption{qARM algorithm for mining frequent itemsets} 
        \label{alg:qARM}
        \hspace*{0.02in} {\bf Input:}  $O_B$, $\mathcal{I}=\{I_0,I_1,\ldots I_{M-1}\}$,  $s_{min}$.\\
        \hspace*{0.02in} {\bf Output:} All the frequent itemsets and their supports.
        \begin{algorithmic}[1]
        \State $k$=1
        \State $\mathcal{C}^{(k)}=\{\{I_0\},\{I_1\},\ldots,\{I_{M-1}\}\}$
        \While{$\mathcal{C}^{(k)} \neq \emptyset$}
                \State $\mathcal{F}^{(k)}=\{\}$
                \State $S^{(k)}=\{\}$
                \State Design $G^{(k)}$ using $O_B$
                \State Create the state of Eq.~\eqref{eq:StartingState} with three quantum registers
                \State Design $G^{(k)}$ using $O_B$
                \State Performing parallel amplitude estimation of $G^{(k)}$ to have state of Eq.~\eqref{eq:pae}
                \State Performing amplitude amplification to find out all the  $k$-itemsets whose supports $\ge s_{min}$  in quantum parallel
                \State Measure the first register and the last register to obtain frequent $k$-itemsets and associated supports,  which are plugged into the set $\mathcal{F}^{(k)}$ and $S^{(k)}$, respectively
                \State Generate $\mathcal{C}^{(k+1)}$ from $\mathcal{F}^{(k)}$ via join and prune
                \State $k=k+1$
        \EndWhile
        \State \Return $\cup_{k}\mathcal{F}^{(k)}$ and $\cup_{k}S^{(k)}$
        \end{algorithmic}
\end{algorithm}

The time complexity of the $k$th iteration of the qARM algorithm is 
$$O\left(k\sqrt{M_c^{(k)}M_f^{(k)}}/\epsilon\right),$$ 
where $M_f^{(k)}$ denotes the number of frequent $k$-itemsets and should be less than $M_c^{(k)}$ and $\epsilon$ is the error of estimating the supports via parallel amplitude estimation (i.e., step 2), whereas the classical counterpart takes  time complexity $O(kM_c^{(k)}/\epsilon^2)$. This means the qARM  algorithm quadratically improves  the time complexity in the dependence of $\epsilon$, and may also attain substantial speedup in the dependence of $M_c^{(k)}$  especially when $M_f^{(k)} \ll M_c^{(k)}$ .

\section{Experimental implementation}\label{sec:experiments}
In this section, we first implement the qARM algorithm for an engineered $2\times 2$ transaction database (i.e., a transaction database with two transactions and two items) on a real IBM quantum computer. To demonstrate qARM for a larger database with size $4\times 4$, we then further implement qARM on a IBM quantum simulator. 

\subsection{Implementation for a $2\times 2$ database on a real IBM quantum computer}

We first consider an engineered tiny database $\mathcal{T}_1=\{T_0,T_1\}$ with two transactions $T_0=\{I_1\},T_1=\{I_0,I_1\}$ and two items $\mathcal{I}=\{I_0,I_1\}$. Similar to the example shown in Fig.~\ref{fig:TDExample}, the corresponding binary matrix of this database is
\begin{align}
        \begin{pmatrix}
                0 & 1\\
                1 & 1
        \end{pmatrix}.
\label{eq:BMTD22}
\end{align}
As a result, the basic quantum oracle $O_B$ would be 
\begin{align}
\ketbra{00}{00}\otimes \mathbb{I}+(\ketbra{01}{01}+\ketbra{10}{10}+\ketbra{11}{11})\otimes X,
\end{align}
where $\mathbb{I}$ is the identity matrix and $X$ is pauli $X$ operation\cite{QCQI10}.

\begin{figure}[!htb]
        \centering
        \includegraphics[width=3in]{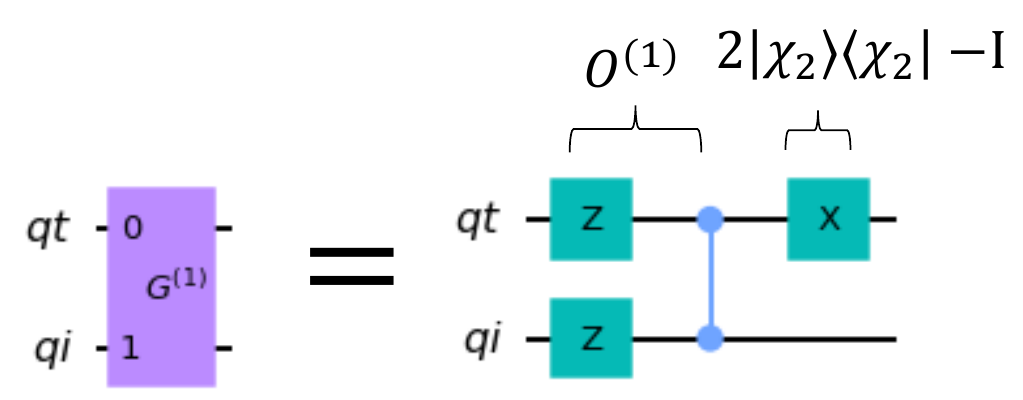}
        \caption{Quantum circuit of $O^{(1)}$ and $G^{(1)}$ for the transaction database $\mathcal{T}_1$. Note that $2\ketbra{\chi_2}{\chi_2}-\mathbb{I}$ is exactly the  Pauli-$X$ operation. The states of qubit $qt$ ($qi$) in the circuit denote different transactions (items) of $\mathcal{T}_1$.}
        \label{fig:qcGrover1TD22}
\end{figure}

Using $O_B$, we construct $O^{(1)}$ that is written as 
\begin{align}
        \ketbra{00}{00}-\ketbra{01}{01}-\ketbra{10}{10}-\ketbra{11}{11},
\end{align}
according to Eq.~\eqref{eq:Ok}. Then we further design the parallel Grover operation $G^{(1)}$ by performing $O^{(1)}$  followed by $2\ketbra{\chi_2}{\chi_2}-\mathbb{I}$, according to Eq.~\eqref{eq:Gk}. Quantum circuits of $O^{(1)}$ and $G^{(1)}$ for $\mathcal{T}_1$ are drawn using Qiskit, an open-source software development kit for quantum computing, in Fig. \ref{fig:qcGrover1TD22}.

Armed with $G^{(1)}$, we design the quantum circuit of qARM for $\mathcal{T}_1$ as depicted in Fig.~\ref{fig:ExpResTD22}, following the steps reviewed in subsection \ref{subsec:qARM}. In this circuit, we however do not introduce the step 3 of qARM, i.e., amplitude amplification, as described in subsection \ref{subsec:qARM}, for two reasons: (1) $\mathcal{T}_1$ has only two items and the computational advantage of amplitude amplification is negligible; (2) involving amplitude amplification would increase the depth of the whole quantum circuit and make qARM not implementable on currently accessible real quantum computers.

\begin{figure}[!htb]
        \centering
        \includegraphics[width=3in]{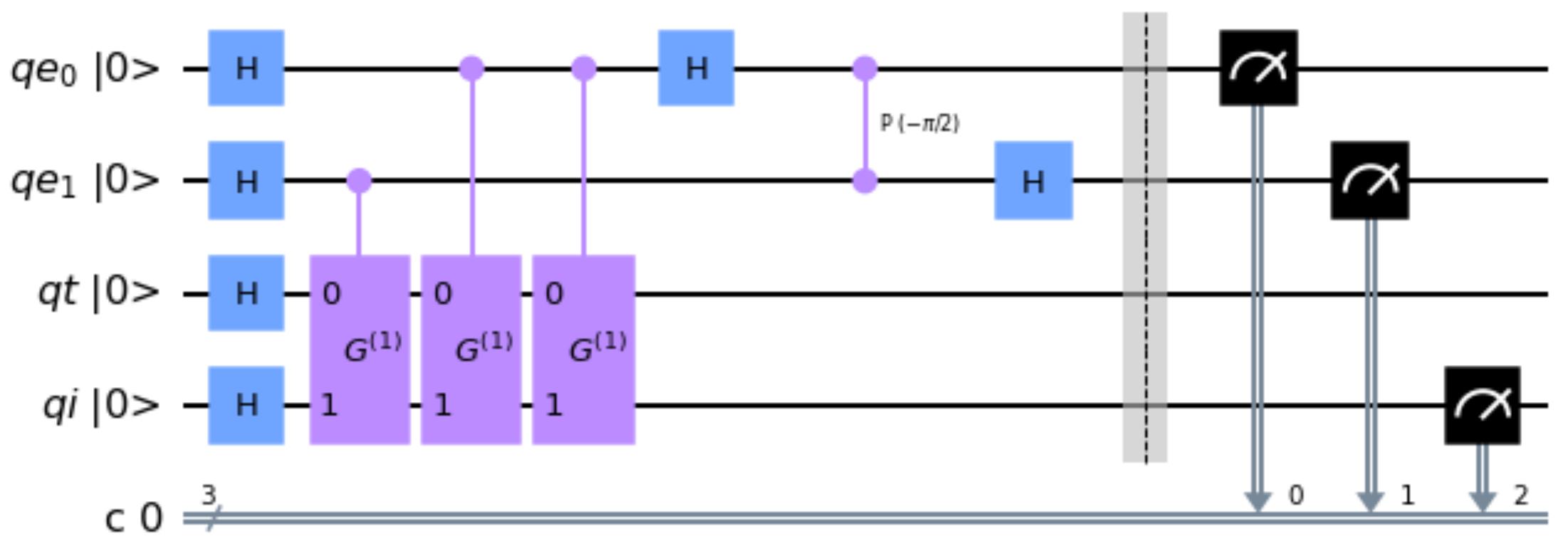}
        \caption{The quantum circuit of qARM for $\mathcal{T}_1$, where the quantum circuit of $G^{(1)}$ is drawn in Fig. \ref{fig:qcGrover1TD22}. As required in \ref{subsec:qARM}, the quantum circuit has three quantum registers and a three-bit classical register. The first quantum register with qubits $qe0$ and $qe1$ aims for estimating eigenvalues of $G^{(1)}$ and the supports of candidate $1$-itemsets, the second quantum register with qubit $qt$ is used to store eigenvectors of $G^{(1)}$ and represent two different transactions, and the last one-qubit quantum register $qi$ is utilized to store two different candidate $1$-itemsets. The final measurement would output the candidate $1$-itemsets and the associated support estimates.}
        \label{fig:qcqARMTD22}
\end{figure}

 We run the quantum circuit of Fig. \ref{fig:qcqARMTD22} on four real five-qubit IBM quantum devices ``ibmq\_quito", ``ibmq\_manila", ``ibmq\_lima" and ``ibmq\_nairobi", as well as on a IBM quantum simulator ``aer\_simulator" that mimics the behavior of an ideal noise-free quantum computer, both with 8192 shots. To see how frequent itemsets are mined in real devices and how far the results in real devices differ from those in the simulator, we concentrate on analyzing the results in the device ``ibmq\_quito" and the analysis applies to other real devices. 

Since the final measurement outcomes are listed in a reverse order in Qiskit, this circuit output three bits labeled by $b_1b_2b_3$, where $b_1=0(1)$ represents the $1$-itemset $\{I_0\}$ ($\{I_1\}$). The decimal of $0.b_2b_3$ corresponds with the estimates of $\theta_j^{(k)}/\pi$ or $1-\theta_j^{(k)}/\pi$ and with support estimates of candidate $1$-itemsets (i.e., $s_j^{(k)}$) according to Eqs.~\eqref{eq:sjk} and \eqref{eq:pae}; since any support ranges from $[0,1]$ and $\theta_j^{(k)} \in [0,\pi/2]$, $\theta_j^{(k)}/\pi$ corresponding with $0.b_2b_3\ge 1/2$ and $1-\theta_j^{(k)}/\pi$ corresponding with $0.b_2b_3\le 1/2$ should occur in pair 
according to Eq.~\eqref{eq:pae}, unless $\theta_j^{(k)}=0$ or $\pi/2$ (i.e., $s_j^{(k)}=0$ or 1). The experimental results of running on the real quantum computer "ibmq\_quito" and on the simulator are presented in Fig. \ref{fig:ExpResTD22:subfig:a}.

Let us first see the simulator results, where there are three measurement outcomes $001$, $011$ and $110$. The outcomes $001$ and $011$ occur in pair and correspond with $\theta_1^{(1)}=\pi/4$, so the support estimate of the candidate 1-itemset $\{I_0\}$ equals 1/2, which coincide with its actual support. The last outcome $110$ occur alone that corresponds with $\theta_2^{(1)}=\pi/2$, which means the support estimate of the candidate 1-itemset $\{I_1\}$ would be one and is consistent with its actual support. 

Now let us see the results of estimating supports on the real quantum computer, and check whether they are consistent with the actual ones. From Fig. \ref{fig:ExpResTD22:subfig:a}, we can see there are eight outcomes $b_1b_2b_3=000,001,\ldots,111$. Since  $\theta_j^{(k)}/\pi$ and $1-\theta_j^{(k)}/\pi$ should occur in pair 
as analyzed above, we take the two outcomes $001$ and $011$ with highest probabilities among the first four outcomes for $b_1=0$ that corresponds to the candidate 1-itemset $\{I_0\}$. These two outcomes are totally consistent with those two on the quantum simulator above, so we will also obtain the support estimate of $\{I_0\}$ equal to 1/2. For $b_1=1$ corresponding with $\{I_1\}$, we have the two outcomes $100$ and $110$ with highest probabilities among the rest four outcomes. However, $100$ and $110$ are not pairwise because they correspond to two different support estimates of $\{I_1\}$, i.e., 0 and 1, so we naturally take the one with a higher probability, i.e., $110$. This outcome implies the support estimate of $\{I_1\}$ is 1 and is consistent with the actual one.

The experimental results of running the qARM circuit (see Fig.~\ref{fig:qcqARMTD22}) both on the real IBM quantum computer "ibmq\_quito" and on the IBM quantum 
simulator "aer\_simulator" demonstrates the correctness 
of qARM for the transaction database $\mathcal{T}_1$. After obtaining the support estimates of both candidate 1-itemsets, one can readily find out the frequent 1-itemsets by comparing the support estimates with the minimum support threshold $s_{min}$. For example, if we set $s_{min}=0.7$, we have only one frequent 1-itemset $\mathcal{F}^{(1)}=\{\{I_1\}\}$.

According to the above analysis for the real quantum device "ibmq\_quito", we can also derive from the subfigures \ref{fig:ExpResTD22:subfig:b} and \ref{fig:ExpResTD22:subfig:c} that the other two real quantum devices ``ibmq\_manila" and ``ibmq\_lima" can also correctly estimate the supports of $\{I_0\}$ and $\{I_1\}$ and successfully dig out the frequent 1-itemset $\mathcal{F}^{(1)}=\{\{I_1\}\}$ as  "ibmq\_quito". However, from the subfigure \ref{fig:ExpResTD22:subfig:d}, we can see that the support of $\{I_0\}$ can be correctly estimated in the real quantum device ``ibmq\_nairobi", but the support of $\{I_1\}$ is wrong estimated to be zero in contrast to its actual support 1. This may be caused by relatively higher noise in ``ibmq\_nairobi". To mitigate errors caused by noise in real quantum computing devices, quantum error mitigation techniques \cite{EBL18,ECBY21} could be a practical solution.

\begin{figure*}[!htb]
        \centering
        \subfloat[]{
                \label{fig:ExpResTD22:subfig:a}
                \includegraphics[width=3.5in]{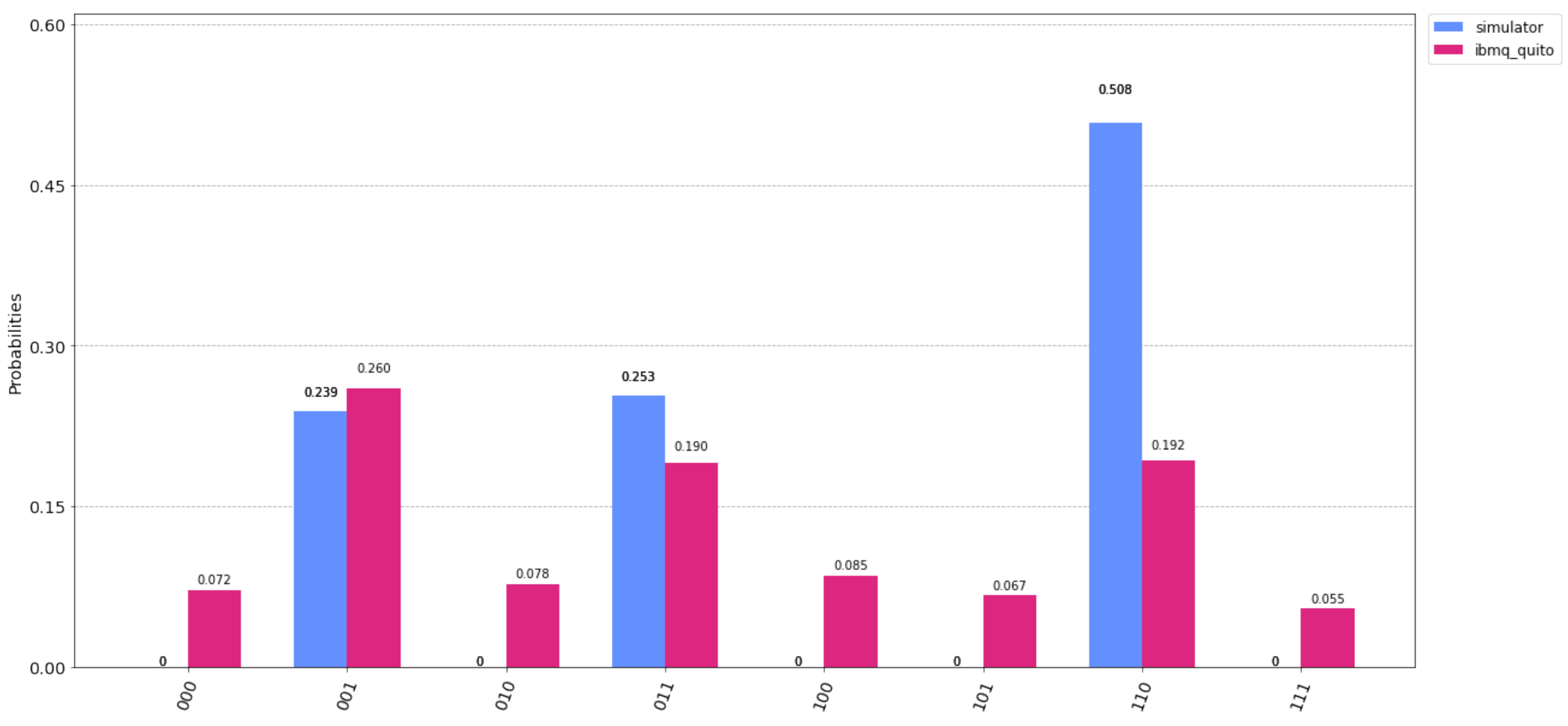}}
        \subfloat[]{
                \label{fig:ExpResTD22:subfig:b}
                \includegraphics[width=3.5in]{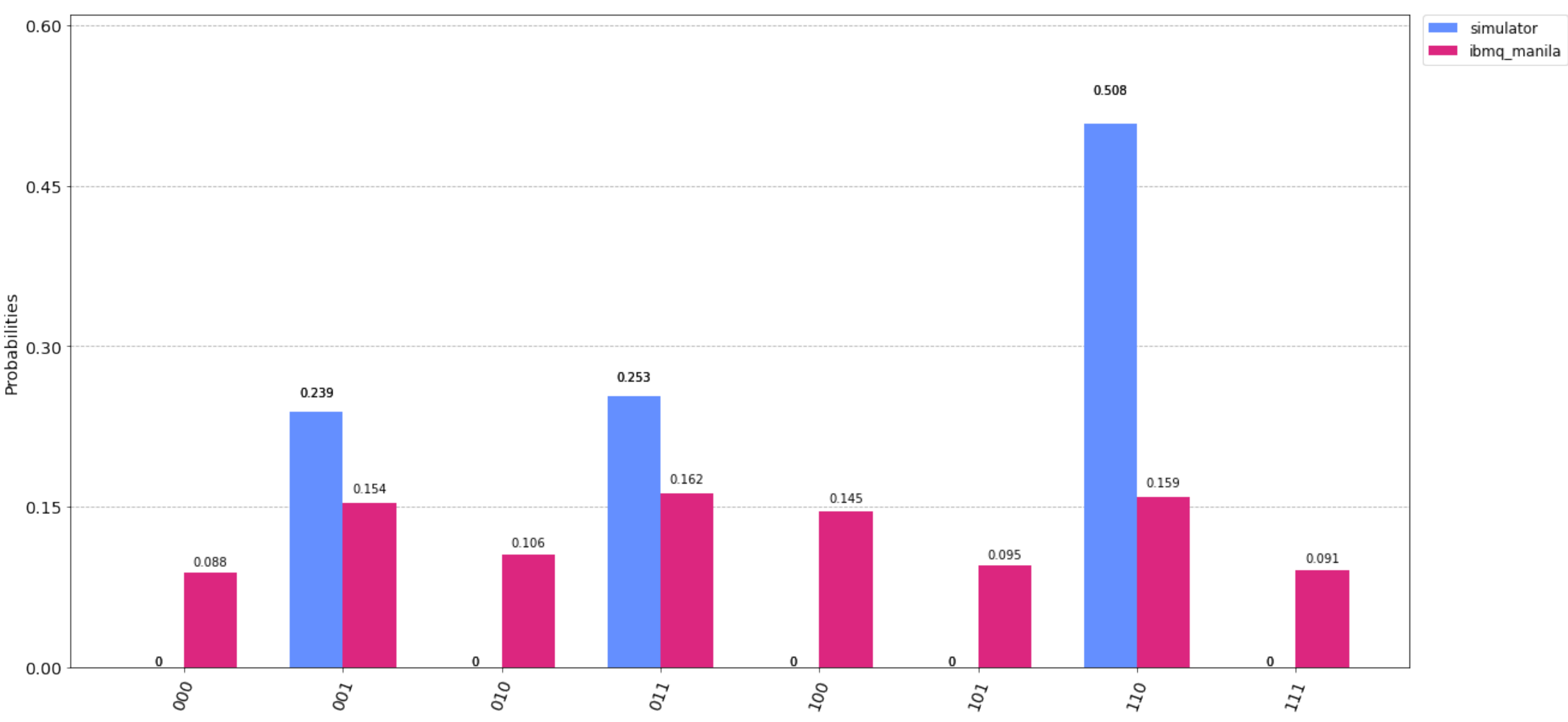}}
                \\
        \subfloat[]{
                \label{fig:ExpResTD22:subfig:c}
                \includegraphics[width=3.5in]{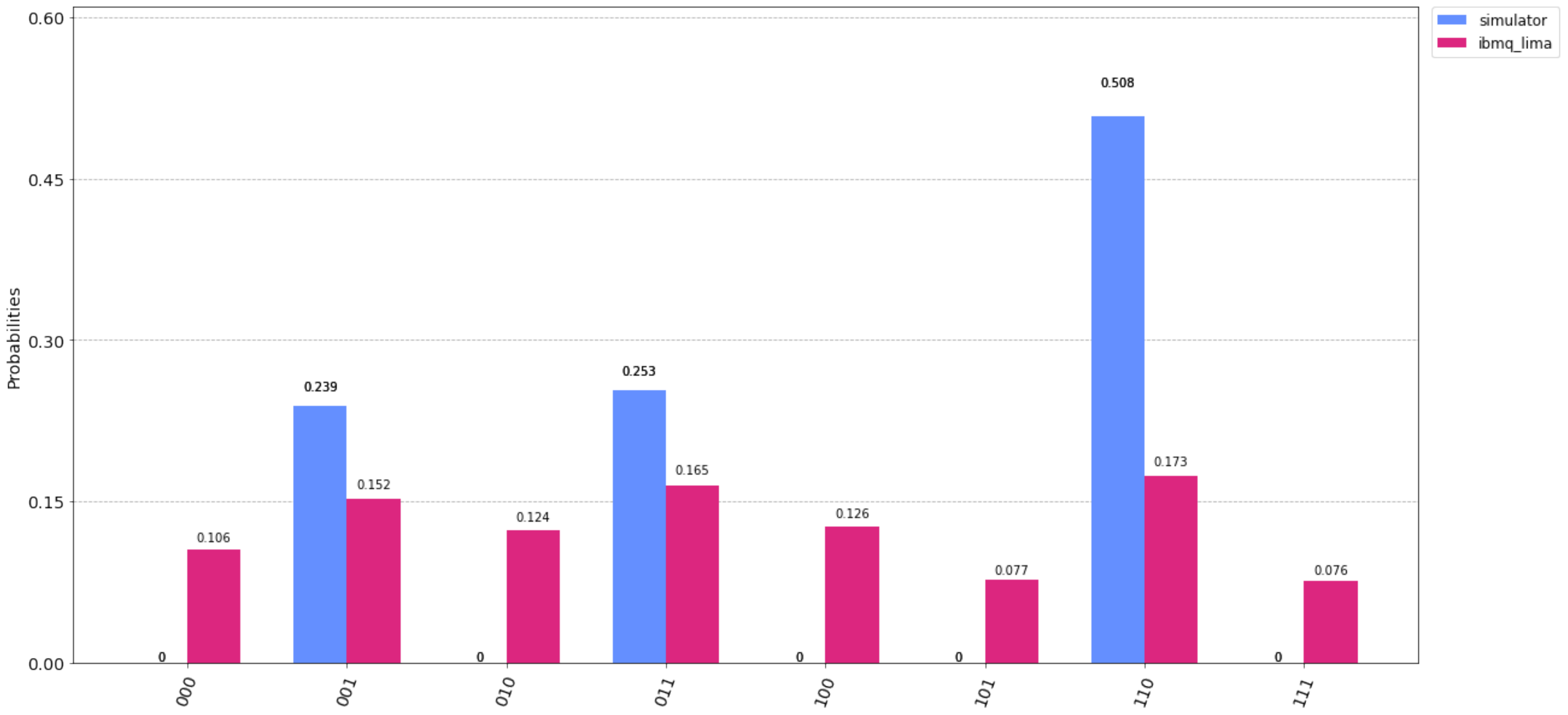}} 
        \subfloat[]{
                \label{fig:ExpResTD22:subfig:d}
                \includegraphics[width=3.5in]{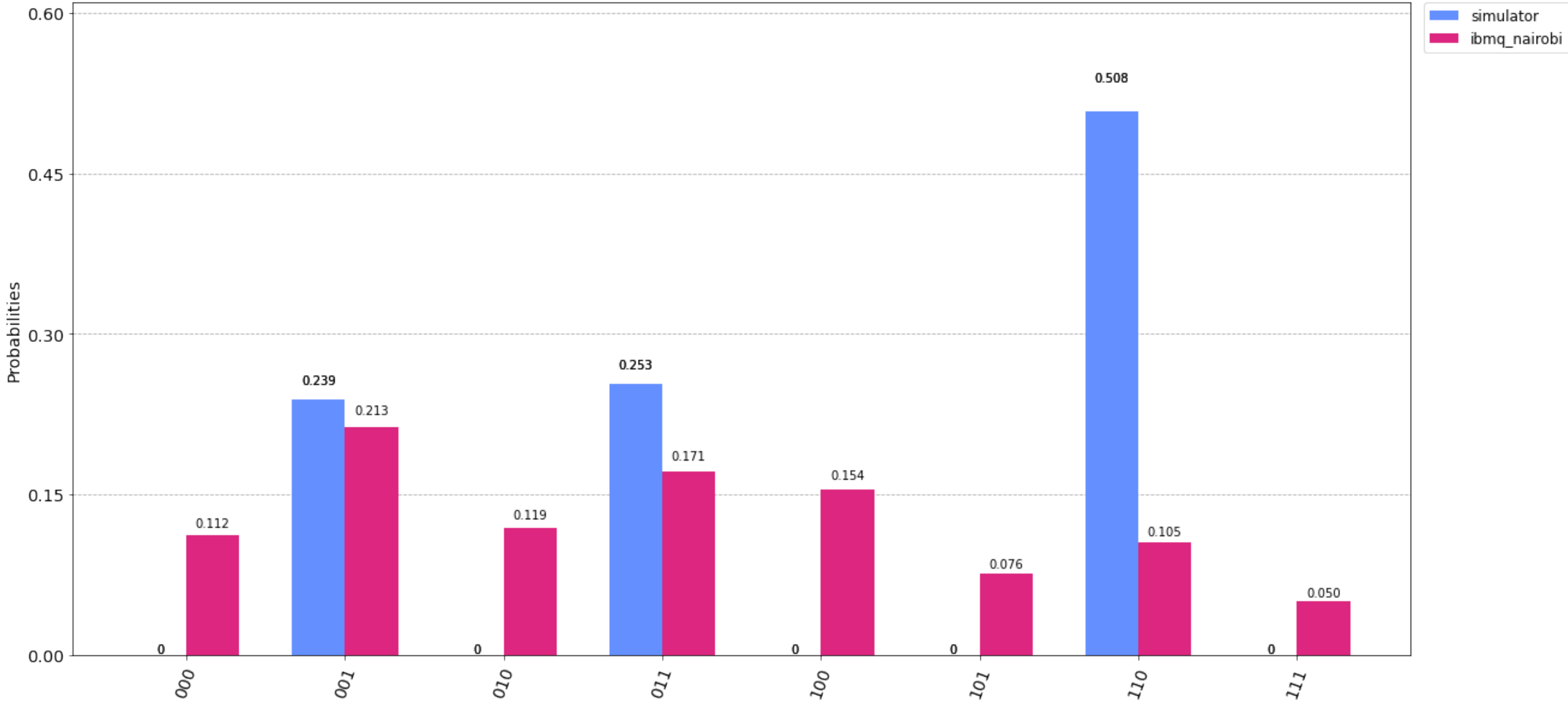}}
        \caption{ Experimental results of running the quantum circuit of Fig. \ref{fig:qcqARMTD22} on four real IBM quantum devices (``ibmq\_quito", ``ibmq\_manila", ``ibmq\_lima" and ``ibmq\_nairobi") and the IBM quantum simulator ``aer\_simulator".}
        \label{fig:ExpResTD22}
\end{figure*}

\subsection{implementation for a $4\times 4$ database on an IBM quantum simulator}

Now we implement qARM for a larger $4\times 4$ transaction database denoted by $\mathcal{T}_2$ presented in Fig.~\ref{fig:TDExample}. The experiment would cover all the steps of qARM listed in subsection \ref{subsec:qARM} including the step (3) of amplitude amplification and the construction of $O^{(k)}$ via $O_B$. Since the qARM circuit for $\mathcal{T}_2$ would require much more qubits and a high depth beyond the limit of currently available open real IBM quantum computers, we choose to implement qARM for $\mathcal{T}_2$ on the IBM quantum simulator ``aer\_simulator''.

\begin{figure}[!htb]
        \centering
        \includegraphics[width=3in]{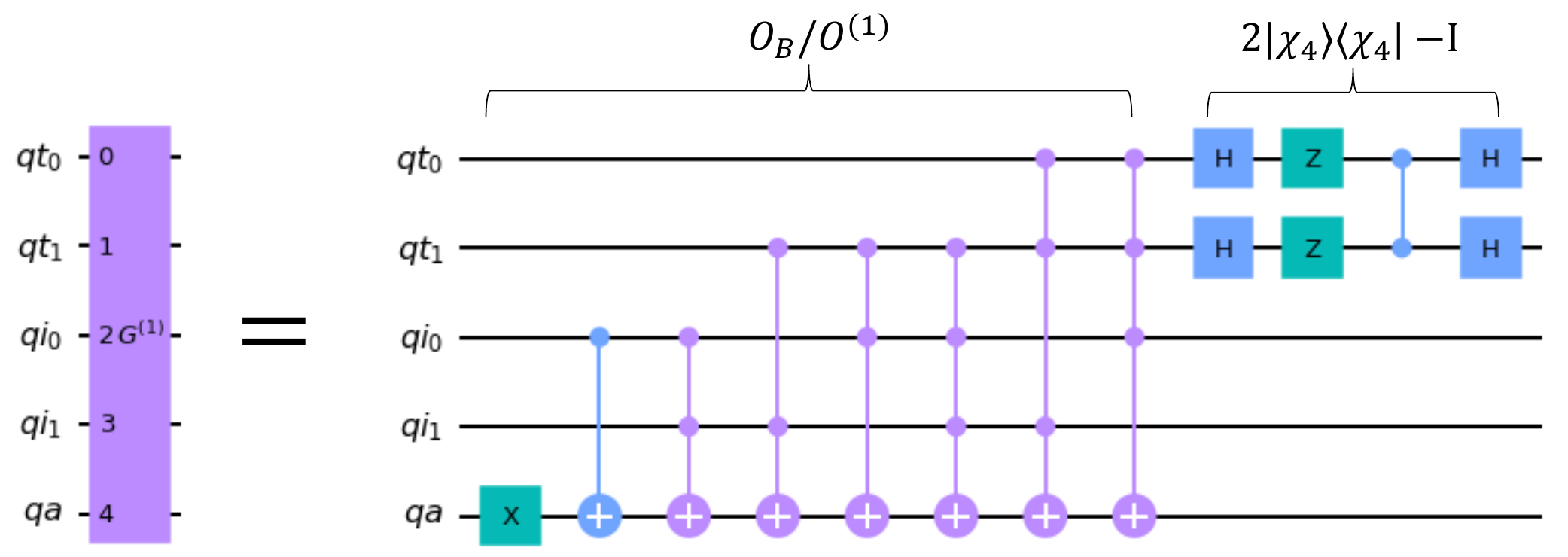}
        \caption{Quantum circuit of $G^{(1)}$ for the transaction database $\mathcal{T}_2$, where $O_B$ (see Eq.~\eqref{eq:OB}) and $O^{(1)}$ (see Eq.~\eqref{eq:O1}) are adopted. The states of qubit $qt_0$ and $qt_1$ in the circuit denote four different transactions of $\mathcal{T}_2$, and those of qubit $qi_0$ and $qi_1$ denote four different items of $\mathcal{T}_2$. The last qubit $qa$ is an ancillary qubit. It is worth noting that if the state of $qa$ is set to be $(\ket{0}-\ket{1})/\sqrt{2}$, the action of $O_B$ on the first four qubits would be just $O^{(1)}$.}
        \label{fig:qcGrover1TD44}
\end{figure}

The construction of $O^{(1)}$ and $G^{(1)}$ together with the basic oracle $O_B$ for $\mathcal{T}_2$ is illustrated in Fig.~\ref{fig:qcGrover1TD44}. An ancillary qubit is introduced to assist the construction. 

\begin{figure}[!htb]
        \centering
        \includegraphics[width=3in]{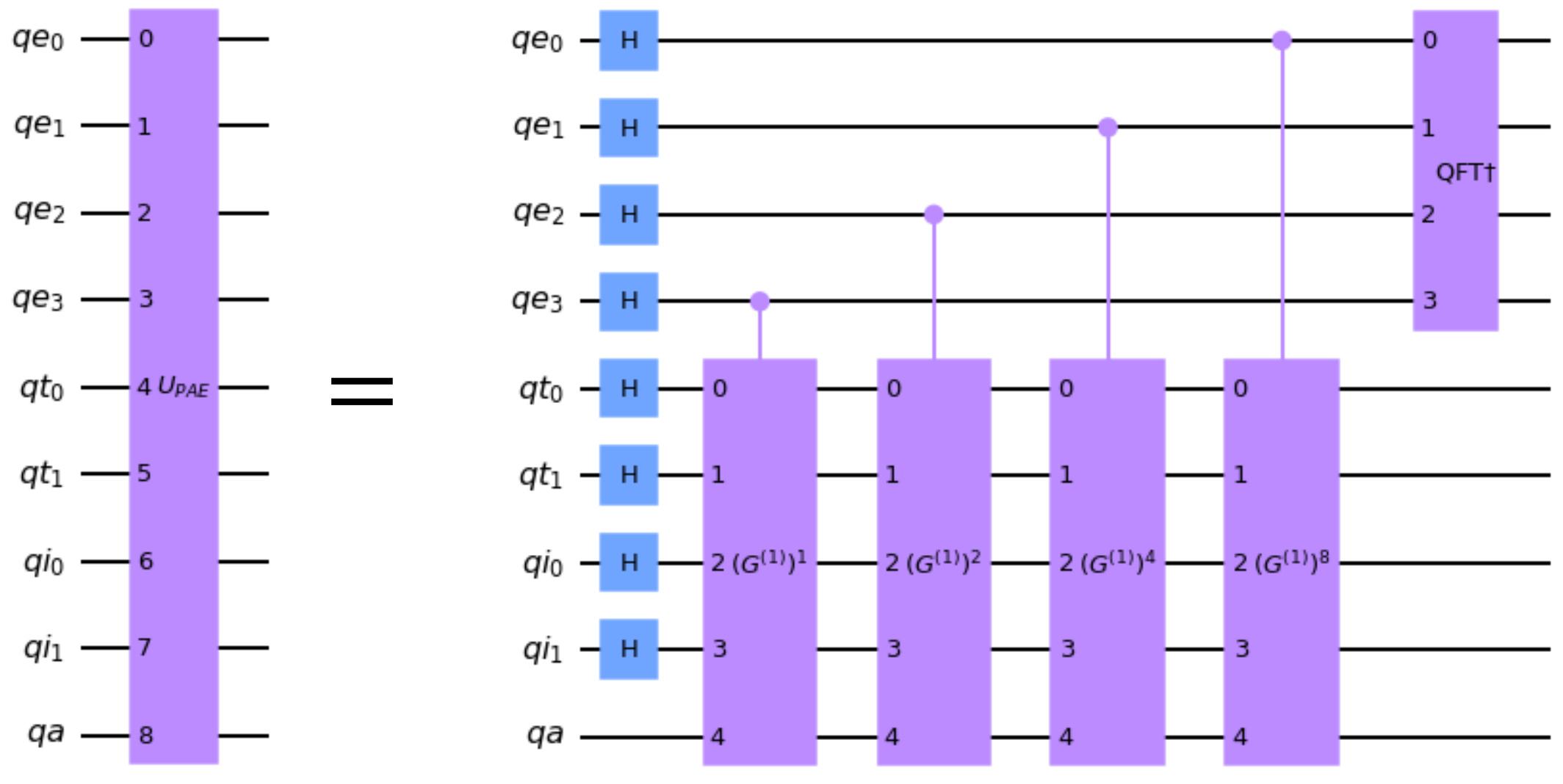}
        \caption{Quantum circuit of parallel amplitude estimation, where the circuit of implementing $G^{(1)}$ is seen in Fig~\ref{fig:qcGrover1TD44}. The circuit includes four quantum registers. The first one contains four qubits $qe_0,qe_1,qe_2,qe_3$ and their states are used to store the eigenvalue estimates of $G^{1}$ and the support estimates of candidate 1-itemsets of $\mathcal{T}_2$. The second one involves two qubits $qt_0, qt_1$ and their states represent four different transactions. The third one also consists of two qubits $qi_0, qi_1$ whose states represents four different 1-itemsets $\{\{I_0\},\{I_1\},\{I_2\},\{I_3\}\}$. The last qubit is an ancillary qubit .Here $\rm QFT\dag$ denotes the inverse quantum Fourier Transformation.} 
        \label{fig:qcPAETD44}
\end{figure}

Taking $G^{(1)}$ as the subroutine, we can further implement the parallel amplitude estimation for step 2 of qARM (described in subsection~\ref{subsec:qARM}), the circuit of which is drawn in Fig.~\ref{fig:qcPAETD44}. This would prepare all the candidate 1-itemsets and their associated support estimates in quantum parallel. 

\begin{figure}[!htb]
        \centering
        \includegraphics[width=3in]{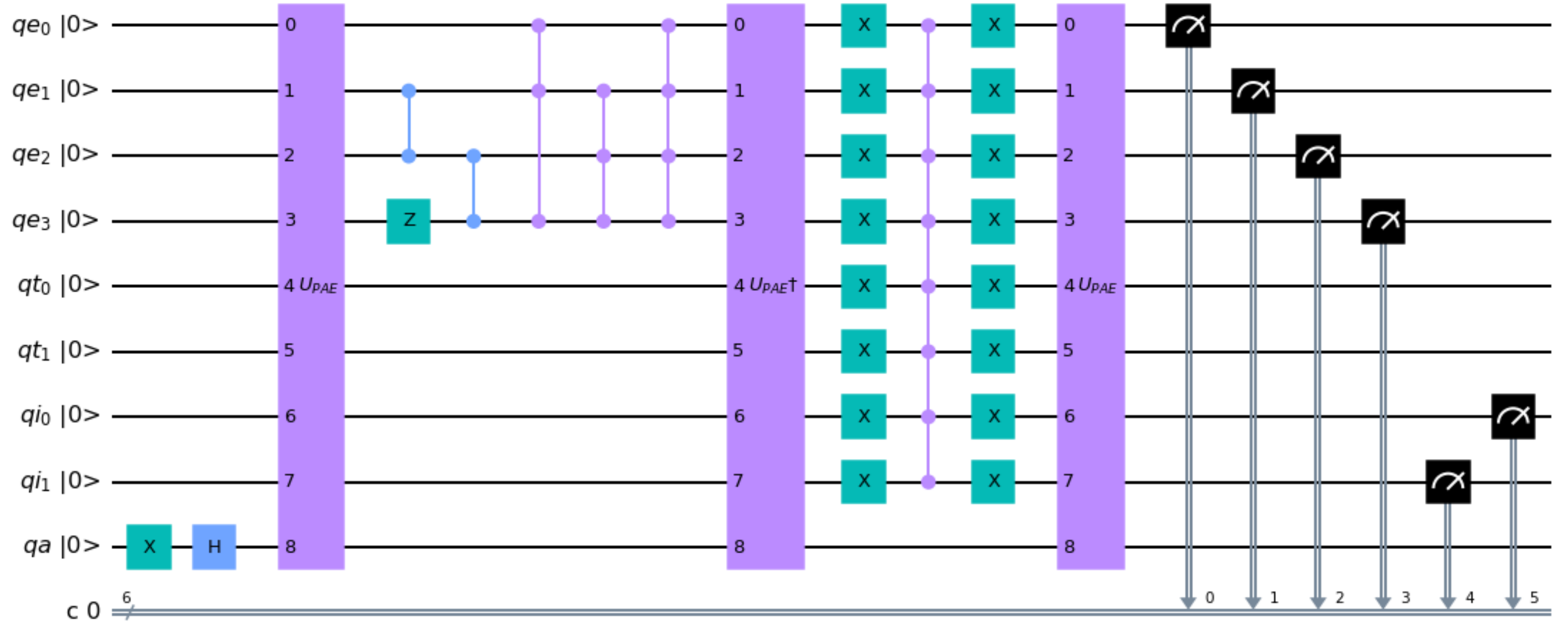}
        \caption{Quantum circuit of qARM for $T_2$ for mining frequent 1-itemsets, where the PAE stands for parallel amplitude estimation and its quantum circuit is illustrated in Fig.~\ref{fig:qcPAETD44}. The quantum circuit goes through the whole steps shown in subsection~\ref{subsec:qARM}: it starts with initializing the last ancillary qubit state  $(\ket{0}-\ket{1})/\sqrt{2}$ by performing $X$ and $H$ gates, subsequently goes to parallel amplitude estimation (i.e. step 1 and 2), then implements the rest gates to fulfill the amplitude amplification part (i.e., step 3 of qARM), and finally performs measurements; the measurement outcomes are stored in the six classical bits. It is worth noting that here we set $s_{min}=0.8$, and thus the phase flip gates between $U_{PAE}$ and $U_{PAE}\dag$, i.e., $O_{s_{min}}$,would flip the phase of the states of the first four qubits $\ket{0110}, \ket{0111}, \ket{1000}, \ket{1001}$, and $\ket{1010}$ according to the Eqs.~\eqref{eq:sjk} and \eqref{eq:pae}.} 
        \label{fig:qcqARMTD44}
\end{figure}

The quantum circuit of qARM for $\mathcal{T}_2$ for mining frequent 1-itemsets is presented in the Fig~\ref{fig:qcqARMTD44}, and the minimum support threshold $s_{min}=0.8$ is set. The circuit mainly includes the parallel amplitude estimation part(steps 1 and 2 described in in subsection~\ref{subsec:qARM}), and the amplitude amplification part (step 3). The whole process of the circuit without measurement can be written as
\begin{align}
U_{PAE}(2\ketbra{0\ldots 0}{0\ldots 0}-\mathbb{I}){U_{PAE}^\dag}O_{s_{min}}{U_{PAE}},
\end{align}
where the unitary operation $U_{PAE}$ stands for the parallel amplitude estimation, and $O_{s_{min}}$ depends on the value of $s_{min}$ and is used to mark  $\ket{\theta_j^{(k)}/\pi}$ and $\ket{1-\theta_j^{(k)}/\pi}$ in Eq.~\eqref{eq:pae} with associated supports greater than or equal to $s_{min}$ according to Eq.~\eqref{eq:sjk} by phase flip.  To construct the circuit of $O_{s_{min}}$, one can also adopt the techniques in Refs. \cite{AAM19,SBBetal21} just as used for implementing $O_B$.  Since here we set $s_{min}=0.8$, $O_{s_{min}}$ would flip the phases of the states $\ket{0110}, \ket{0111}, \ket{1000}, \ket{1001}$, and $\ket{1010}$. The final measurement output takes six bits denoted by $b_1b_2b_3b_4b_5b_6$, and the values of $b_1b_2$ represents four different candidate 1-itemsets  $\{\{I_0\},\{I_1\},\{I_2\},\{I_3\}\}$, and $b_3b_4b_5b_6$
are taken to estimate the associated supports. 

\begin{figure*}[!t]
        \centering
        \includegraphics[width=6.5in]{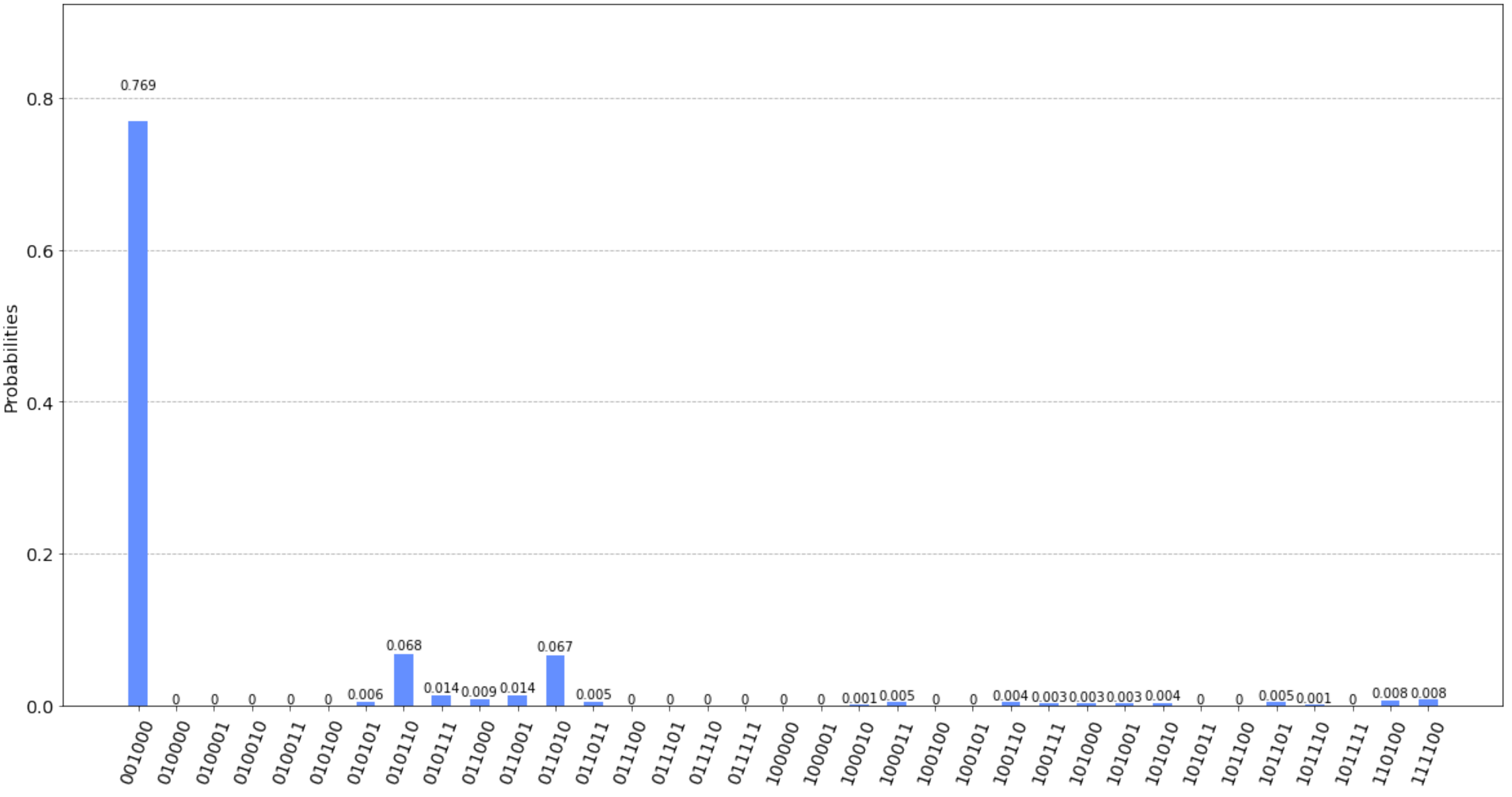}
        \caption{Histogram of the measurement outcomes of running the quantum qARM circuit of Fig.~\ref{fig:qcqARMTD44} for $1024\times 128$ shots on the IBM quantum simulator ``aer\_simulator''.}
        \label{fig:ExpResTD44}
\end{figure*}

The above circuit is run on the IBM quantum simulator ``aer\_simulator'' with $1024\times128$ shots, and the histogram of measurement outcomes is depicted in Fig.~\ref{fig:ExpResTD44}. From this figure, it is easy to see that the outcome $0010000$ has a dominantly high probability, which implies only the candidate 1-itemset $\{I_{0}\}$ is frequent. This is perfectly consistent with the actual result due to the fact that only $\{I_{0}\}$ has support one that is greater than  $s_{min}=0.8$, which can be easily seen from Fig.~\ref{fig:TDExample}. Then whole quantum ARM algorithm ends with only one frequent itemset, i.e., $\{I_{0}\}$.  

\begin{figure}[!htb]
        \centering
        \includegraphics[width=3in]{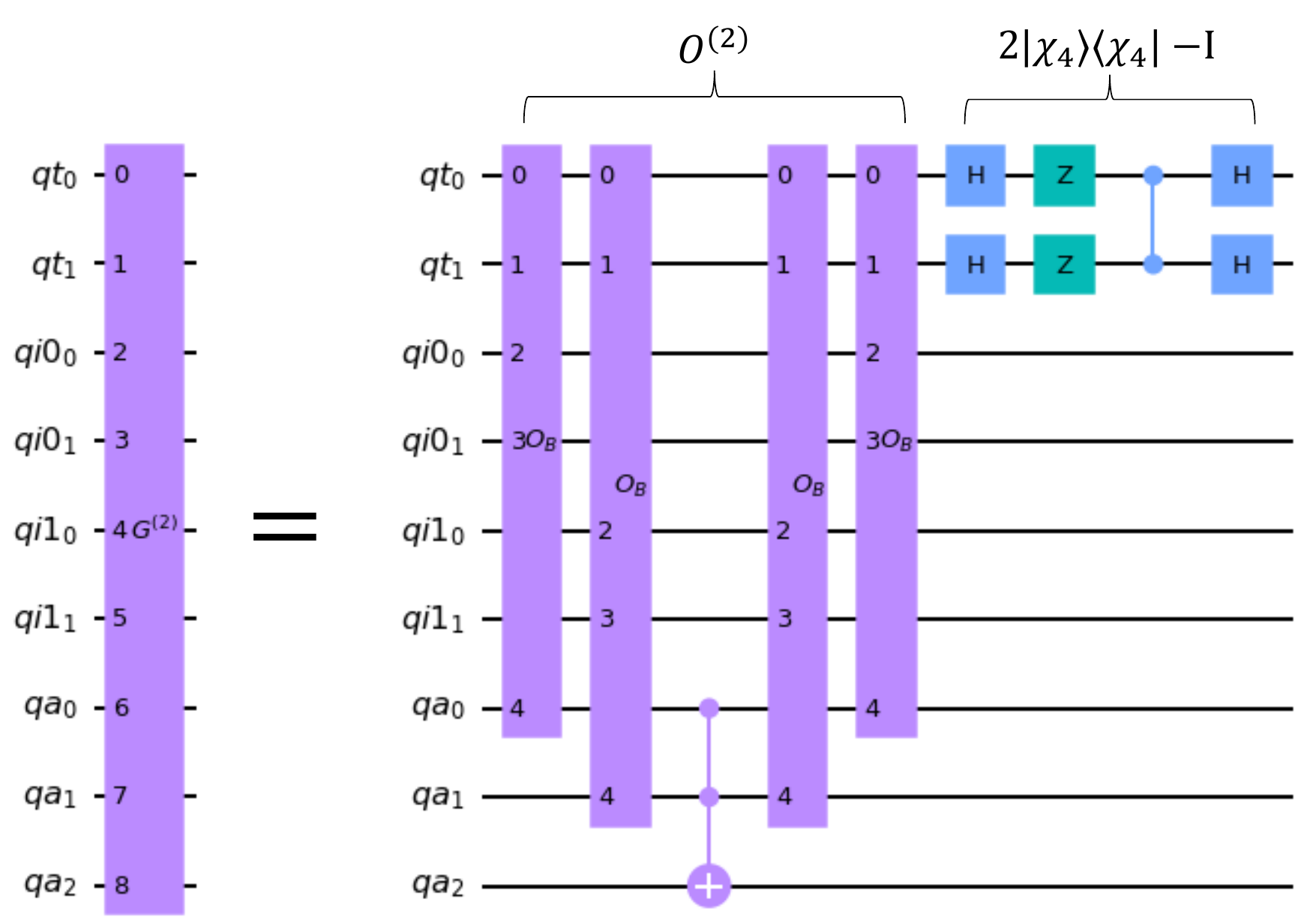}
        \caption{Quantum circuit of $G^{(2)}$ for the transaction database $\mathcal{T}_2$, where $O_B$ (see Eq.~\eqref{eq:OB} and Fig. \ref{fig:qcGrover1TD44}) is taken to construct $O^{(2)}$ (see Eq.~\eqref{eq:Ok}). In the circuit, the states of the first two qubits $qt_0$ and $qt_1$ represent four different transactions of $\mathcal{T}_2$, and those of qubits $qi0_0$ and $qi0_1$ ($qi1_0$ and $qi1_1$) denote the first (second) item of any candidate 2-itemsets of $\mathcal{T}_2$. The last three qubits $qa_0$, $qa_1$ and $qa_2$ are ancillary qubits. }
        \label{fig:qcGrover2TD44}
\end{figure}

Since the above experiments only address estimating the supports of candidate 1-itemsest and mining frequent 1-itemsets, one may wonder how to experimentally estimate the supports of 2-itemsets. Let us still take the above example presented in Fig.~\ref{fig:TDExample}, and consider estimating the supports of two 2-itemsets  $\{I_0,I_2\}$ and $\{I_1,I_3\}$ whose actual supports are 1/4 and 1/2 respectively. In the first place, following steps of qARM presented in Ref. \cite{QARM16}, we design the quantum circuit of the Grover operation $G^{(2)}$ for $\mathcal{T}_2$ shown in Fig. \ref{fig:qcGrover2TD44}. Different from the quantum circuit of $G^{(1)}$ with only one ancillary qubit, the quantum circuit of $G^{(2)}$ introduce three ancillary qubits.

\begin{figure}[!htb]
        \centering
        \includegraphics[width=3in]{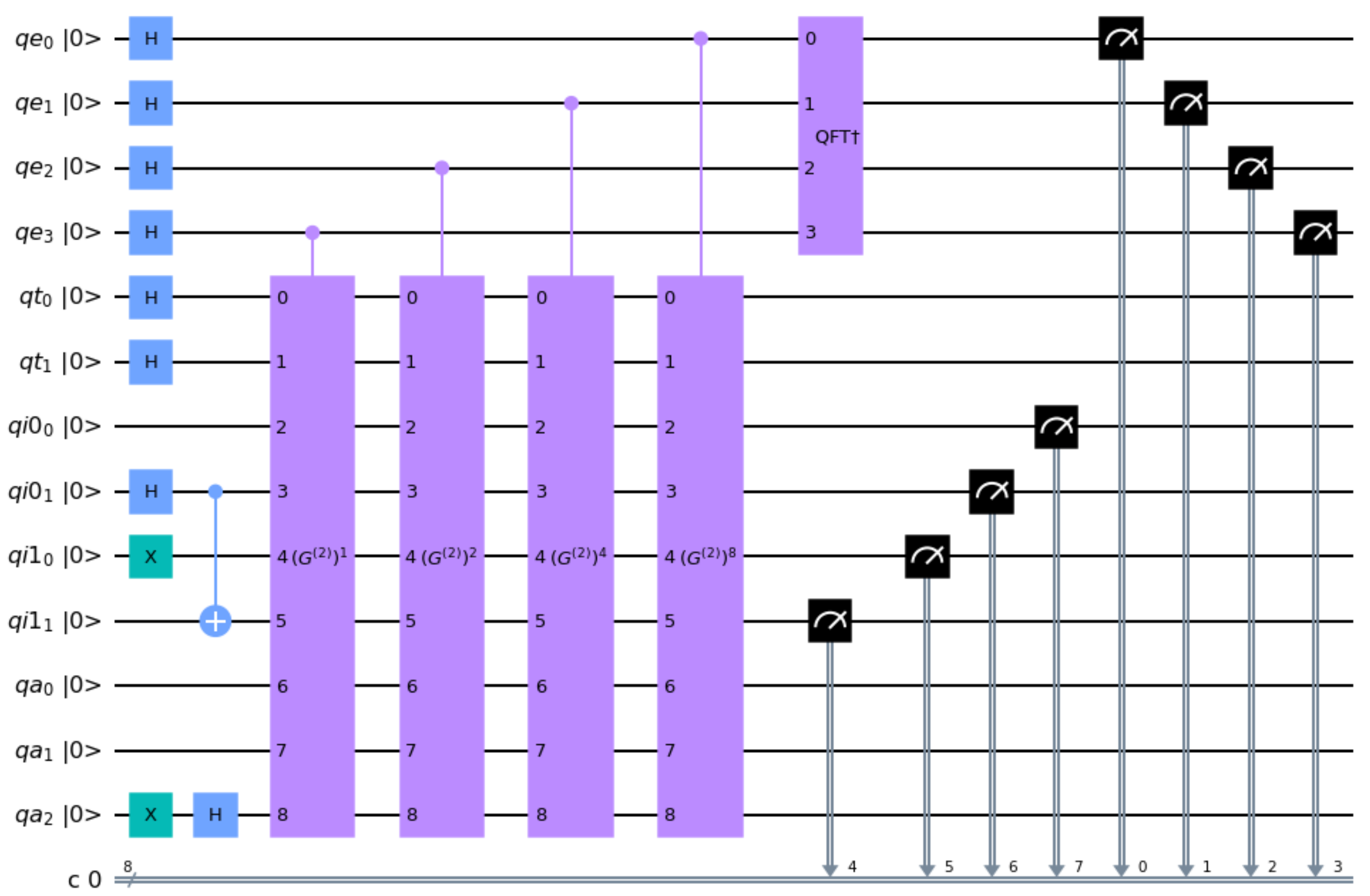}
        \caption{Quantum circuit of qARM for estimating supports of the two  2-itemsets $\{I_0,I_2\}$ and $\{I_1,I_3\}$ of $\mathcal{T}_2$, where the quantum circuit of $G^{(2)}$ is depicted in Fig~\ref{fig:qcGrover2TD44}. }
        \label{fig:qcPAE2TD44}
\end{figure}

Using $G^{(2)}$, we further design the quantum circuit of parallel quantum amplitude estimation for estimating the supports of the two 2-itemsets $\{I_0,I_2\}$ and $\{I_1,I_3\}$, which is illustrated in Fig.~\ref{fig:qcPAE2TD44}. In the circuit, the final measurement outputs eight bits, where the first four bits represent different 2-itemsets and the last four bits gives the corresponding support estimates.

\begin{figure*}[!htb]
        \centering
        \includegraphics[width=6.5in]{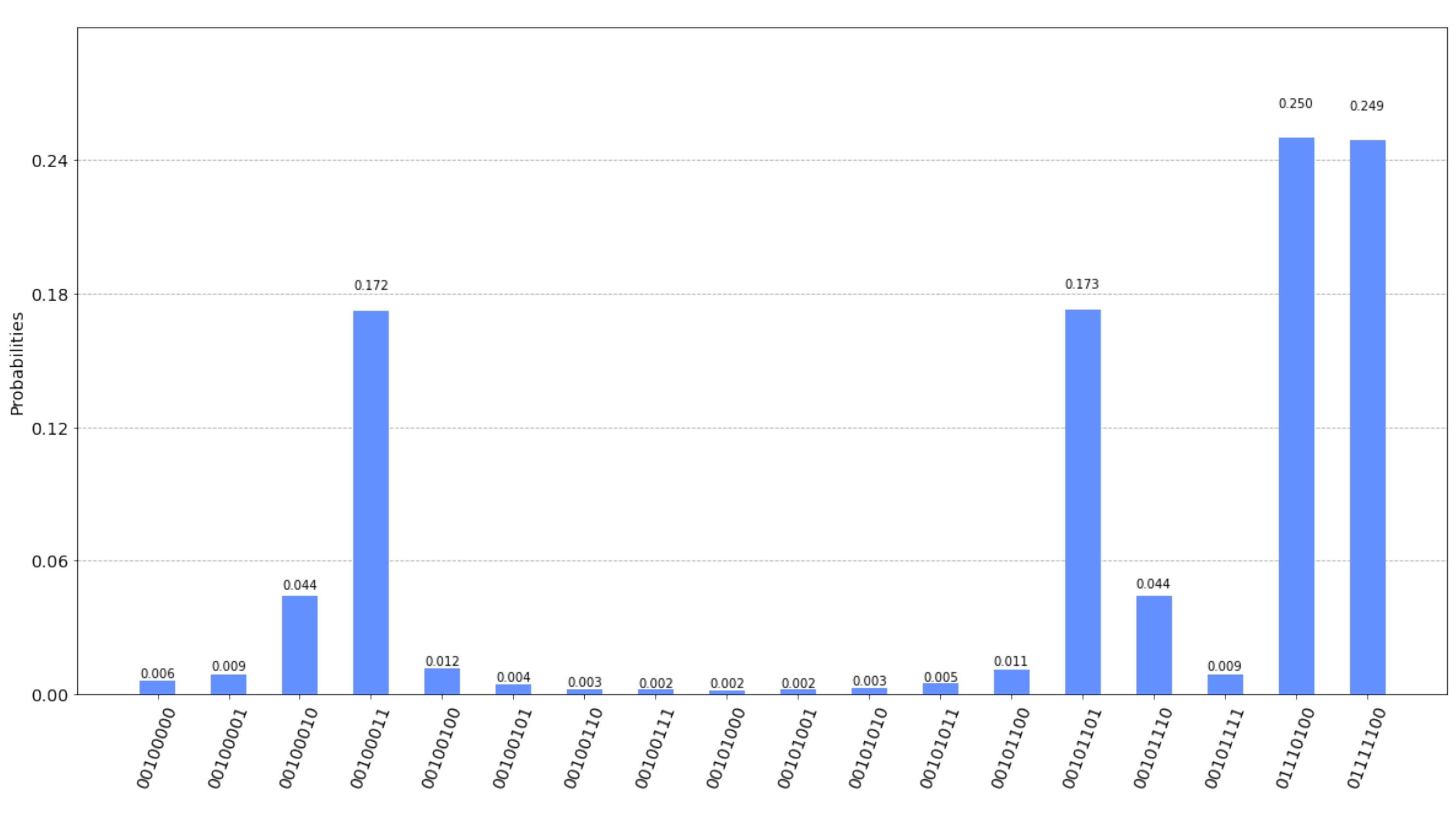}
        \caption{Histogram of measurement results after running the quantum circuit of Fig.~\ref{fig:qcPAE2TD44} for $1024\times 128$ shots on the IBM quantum simulator ``aer\_simulator''.}
        \label{fig:ExpRes2TD44}
\end{figure*}

We run the quantum circuit of Fig.~\ref{fig:qcPAE2TD44} on the IBM quantum simulator ``aer\_simulator'' with $1024\times 128$ shots, and the results are presented in the Fig.~\ref{fig:ExpRes2TD44}. From this figure, we can see the support of $\{I_0,I_2\}$ is estimated as 0.691, which is closer to the actual support 3/4 considering its support can only be one of $\{0,1/4,2/4,3/4,1\}$, so its support estimate would be deemed to be equal to the actual value of 3/4. Similarly, the support of $\{I_1,I_3\}$ is estimated as 1/2, equal to its actual support. The above results show that we are able to correctly estimate the supports of 2-itemsest and thus mine frequent 2-itemsest on a noiseless quantum computer.

\section{Conclusions}

In this paper, we have provided the first proof-of-principle demonstration of qARM via IBM quantum platform. In the first place, we design the quantum circuit of qARM for a tiny $2\times 2$ transaction database and runs it on four real IBM quantum computing devices and on a noise-free IBM quantum simulator. For a larger $4\times 4$ transaction database which cannot be used for experimental implementation to generate reasonable results on real devices due to depth limit, we design the quantum circuit of qARM for this database and instead execute it on the simulator. The experimental results for both databases show that the frequent itemsets hidden in these two databases can be successfully mined via qARM on most real devices and simulator as desired, showing the correctness and feasibility of qARM. Moreover, our experimental implementation provides prototypes for implementing qARM for larger databases on near-term noisy intermediate-scale quantum ( NISQ) computers and future universal fault-tolerant quantum computers, and may inspire more implementations of other quantum algorithms for machine learning and data mining.

 Meanwhile, our experimental implementation also reveals the hardness of currently accessible real quantum computing devices for  implementing qARM for even tiny databases. To make qARM for larger databases more effectively implementable on real devices, practical quantum error mitigation techniques \cite{AAM19,SBBetal21} could be a solution.

\section{Acknowledgments}
This work is supported by the National Natural
Science Foundation of China (Grant Nos. 62006105, 61976024), and the Jiangxi Provincial Natural
Science Foundation (Grant No. 20202BABL212004).

 
\vspace{11pt}


\begin{IEEEbiography}[{\includegraphics[width=1in,height=1.25in,clip,keepaspectratio]{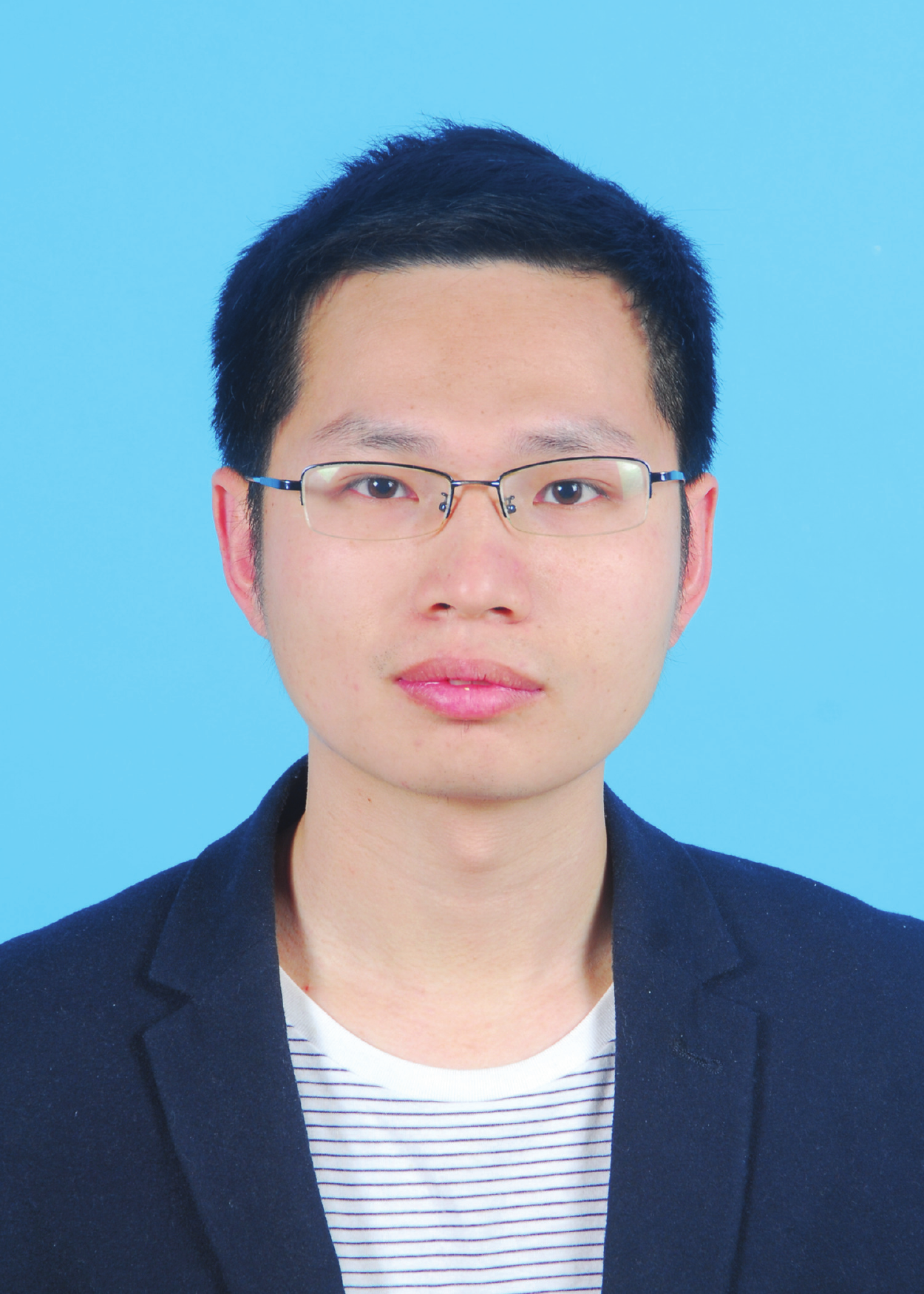}}]{Chao-Hua Yu} received the Ph.D. degree in cryptography at Beijing university of Posts and Telecommunication, Beijing, China, in 2019.
He is currently a lecturer with the School of Information Management, Jiangxi University of Finance and Economics, Nanchang, China. His research interests include quantum algorithms for machine learning and data mining, and quantum cryptography.
\end{IEEEbiography}



\end{document}